\documentclass[twocolumn]{aastex63}

\usepackage{lineno}
\usepackage[figuresright]{rotating}

\received{***, ****}
\revised{***, ****}
\accepted{***, ****}
\submitjournal{ApJL}

\begin{document}
\title{Alpha-Proton Differential Flow of A Coronal Mass Ejection at 15 Solar Radii}

\author{Xuechao Zhang}
\affiliation{Institute of Frontier and Interdisciplinary Science, Shandong University, Qingdao, Shandong 266237, China}
\affiliation{Shandong Provincial Key Laboratory of Optical Astronomy and Solar-Terrestrial Environment, and Institute of Space Sciences, Shandong University, Weihai, Shandong 264209, China}

\correspondingauthor{Hongqiang Song}
\email{hqsong@sdu.edu.cn}

\author{Hongqiang Song}
\affiliation{Shandong Provincial Key Laboratory of Optical Astronomy and Solar-Terrestrial Environment, and Institute of Space Sciences, Shandong University, Weihai, Shandong 264209, China}
\affiliation{State Key Laboratory of Space Weather, National Space Science Center, Chinese Academy of Sciences, Beijing 100190, China}

\author{Xiaoqian Wang}
\affiliation{Shandong Provincial Key Laboratory of Optical Astronomy and Solar-Terrestrial Environment, and Institute of Space Sciences, Shandong University, Weihai, Shandong 264209, China}

\author{Leping Li}
\affiliation{National Astronomical Observatories, Chinese Academy of Sciences, Beijing, 100101, China}

\author{Hui Fu}
\affiliation{Shandong Provincial Key Laboratory of Optical Astronomy and Solar-Terrestrial Environment, and Institute of Space Sciences, Shandong University, Weihai, Shandong 264209, China}

\author{Rui Wang}
\affiliation{Shandong Provincial Key Laboratory of Optical Astronomy and Solar-Terrestrial Environment, and Institute of Space Sciences, Shandong University, Weihai, Shandong 264209, China}

\author{Yao Chen}
\affiliation{Shandong Provincial Key Laboratory of Optical Astronomy and Solar-Terrestrial Environment, and Institute of Space Sciences, Shandong University, Weihai, Shandong 264209, China}
\affiliation{Institute of Frontier and Interdisciplinary Science, Shandong University, Qingdao, Shandong 266237, China}

\begin{abstract}
Alpha-proton differential flow ($V_{\alpha p}$) of coronal mass ejections (CMEs) and solar wind from the Sun to 1 au and beyond could influence the instantaneous correspondence of absolute abundances of alpha particles (He$^{2+}$/H$^{+}$) between solar corona and interplanetary space as the abundance of a coronal source can vary with time. Previous studies based on Ulysses and Helios showed that $V_{\alpha p}$ is negligible within CMEs from 5 to 0.3 au, similar to slow solar wind ($<$ 400 km s$^{-1}$). However, recent new observations using Parker Solar Probe (PSP) revealed that the $V_{\alpha p}$ of slow wind increases to $\sim$60 km s$^{-1}$ inside 0.1 au. It is significant to answer whether the $V_{\alpha p}$ of CMEs exhibits the similar behavior near the Sun. In this Letter, we report the $V_{\alpha p}$ of a CME measured by PSP at $\sim$15 $R_\odot$ for the first time, which demonstrates that the $V_{\alpha p}$ of CMEs is obvious and complex inside 0.1 au while keeps lower than the local Alfv\'{e}n speed. A very interesting point is that the same one CME duration can be divided into A and B intervals clearly with Coulomb number below and beyond 0.5, respectively. The means of $V_{\alpha p}$ and alpha-to-proton temperature ratios of interval A (B) is 96.52 (21.96) km s$^{-1}$ and 7.65 (2.23), respectively. This directly illustrates that Coulomb collisions play an important role in reducing the non-equilibrium features of CMEs. Our study indicates that the absolute elemental abundances of CMEs also might vary during their propagation.
\end{abstract}


\section{Introduction}
Coronal mass ejections \citep[CMEs,][]{chenpengfei11,webb12}, the most energetic eruption in the solar atmosphere, are often correlated with filament eruption \citep{gopalswamy03,song13,song18a,chengxin20,wangxinyue23}. Researchers have used the in situ charge states of heavy ions and the abundances of heavy elements relative to oxygen (i.e., relative abundances) near 1 au to investigate the eruption process of CMEs \citep{song16,wangwensi17} and the origin of filament \citep{song17a,lepri21} in the corona, see \cite{song20b} for a review.

As the elemental abundance of a coronal source can vary with time \citep{widing01,baker13}, the instantaneous correspondence of relative abundances of heavy elements between solar corona and interplanetary space requires no differential flow among various heavy ions as an essential precondition \citep{zhangxuechao24}, which is met within both CMEs \citep{zhangxuechao24} and solar wind \citep{vonSteiger95,vonSteiger06,berger11}. Likewise, to correlate the instantaneous abundances of alpha particles relative to hydrogen (i.e., absolute abundances), it requires no alpha-proton differential flow ($V_{\alpha p}$) within CMEs and solar wind during their propagation to 1 au and beyond.

The $V_{\alpha p}$ of solar wind was first reported near 1 au in 1970 \citep{robbins70,formisano70}. Subsequently, observations of Helios and Ulysses demonstrated that the $V_{\alpha p}$ in fast wind decreases with increasing heliocentric distance \citep{reisenfeld01}. For instance, the $V_{\alpha p}$ of fast wind ($>$ 500 km s$^{-1}$) decreases from $\sim$150 km s$^{-1}$ at 0.3 au to $\sim$40 km s$^{-1}$ at 1 au \citep{marsch82b}, with its magnitude comparable to or lower than the local Alfv\'{e}n speed ($V_{A}$) and direction parallel to the local interplanetary magnetic field \citep[IMF;][]{marsch82b,neugebauer94,neugebauer96}. In the meanwhile, alphas are usually hotter than protons in solar wind \citep{kasper08,maruca13}. Recent statistical study showed that the histogram of alpha-to-proton temperature ratio (T$_{\alpha}/$T$_{p}$) has a maximum at T$_{\alpha}/$T$_{p}$=4 for fast wind near 1 au \citep{durovcova17}.

The above observations imply the preferential acceleration and heating of alphas. Researchers have proposed different mechanisms to explain them, such as the stochastic heating induced by low-frequency Alfv\'{e}n turbulence \citep{chandran10} and the resonant absorption of ion-cyclotron wave \citep{kasper13}. In the meantime, the alpha-proton instability \citep{gary00} and Coulomb collisions \citep{kasper08} reduce the $V_{\alpha p}$ and $T_{\alpha}/T_{p}$ \citep{chhiber16,durovcova17}.

Contrary to the fast wind, the $V_{\alpha p}$ of slow wind ($<$ 400 km s$^{-1}$) and CMEs keeps negligible from 0.3 to 1 au and beyond \citep{marsch82b,liuying06,durovcova17,zhangxuechao24}. However, recent new observations of Parker Solar Probe \citep[PSP;][]{fox16} revealed that the radial dependence of $V_{\alpha p}$ holds inside 0.3 au and the $V_{\alpha p}$ of slow wind reaches $\sim$60 km s$^{-1}$ inside 0.1 au \citep{mostafavi22}. This means that the instantaneous observations of absolute abundances of alphas of both slow and fast winds at 1 au might not represent their corresponding abundances in the corona, and raises an important question: whether CMEs have nonnegligible $V_{\alpha p}$ or not near the Sun. This is our motivation to conduct the current study. In Section 2, we introduce the instruments and methods. The observations and results are displayed in Section 3, which is followed by a summary and discussion in the final section.

\section{Instruments and Methods}
\subsection{Instruments}
The PSP was launched in 2018 to unravel the mysteries of the Sun at a closer distance than any previous spacecraft \citep{fox16}. In this study, we use the data provided by Solar Wind Electrons Alphas and Protons (SWEAP) that measures ion and electron distribution functions and corresponding moments \citep{kasper16}, and FIELDS that measures magnetic and electric fields \citep{bale16}. All data are averaged to a cadence of 10 seconds. The SWEAP consists of two ion sensors, the Solar Probe Analyzer for Ions (SPAN-I) and the Solar Probe Cup. Here the SPAN-I is used, which consists of a ram-facing electrostatic analyzer and time-of-flight section, enabling to distinguish ions with distinct masses, such as protons and alphas, according to their mass-per-charge ratios.

The Atmospheric Imaging Assembly \citep[AIA;][]{lemen12} aboard Solar Dynamics Observatory \citep[SDO;][]{pesnell12}, as well as the Extreme Ultraviolet Imager (EUVI) and white-light coronagraphs (COR2-A) \citep{howard08} aboard Solar Terrestrial Relations Observatory \citep[STEREO;][]{kaiser05} provide the remote sensing observations.

\subsection{Methods}
Various methods have been adopted to calculate the differential flow. Some previous studies \citep{asbridge76,neugebauer94,kasper08} used Equation (1) to get the $V_{\alpha p}$, which calculates the speed difference directly without considering the direction. Meanwhile, some studies \citep{reisenfeld01,fuhui18} calculated the $V_{\alpha p}$ along the local magnetic field using Equation (2) as the differential flow is parallel to the field, in which $V_{r\alpha}$ and $V_{rp}$ are the radial speed of alphas and protons, respectively. The $\theta$ denotes the angle between the field direction and the radial vector. Recently, many studies \citep{durovcova17,mostafavi22,mostafavi24,ranhao24} calculated the $V_{\alpha p}$ through Equation (3), which is used in the current study as protons and alphas usually flow along different directions. It does not mean no differential flow when Equation (1) equals to zero \citep{durovcova17}. Note that the bold and nonbold symbols represent the vectors and vector magnitudes, respectively.

\begin{equation}
\rm V_{\alpha p}=|\mathbf{V_{\alpha}}|-|\mathbf{V_{p}}|
\end{equation}

\begin{equation}
\rm V_{\alpha p}=\frac{V_{r\alpha}-V_{rp}}{cos\theta}
\end{equation}

\begin{equation}
\rm V_{\alpha p}=sign(\left|\mathbf{{V}_{\alpha}}\right|-\left|\mathbf{{V}_{p}}\right|) \cdot \left|\mathbf{{V}_{\alpha}} - \mathbf{{V}_{p}}\right|
\end{equation}

To compare the $V_{\alpha p}$ with the local $V_{A}$, we compute the $V_{A}$ using Equation (4)
\begin{equation}
\rm V_A=\frac{B}{\sqrt{\mu_0\left(N_p m_p+N_\alpha m_\alpha\right)}}
\end{equation}
where B and $\mu_{0}$ are the magnetic-field magnitude and vacuum permeability, respectively, $N_{p}$ ($m_{p}$) and $N_{\alpha}$ ($m_{\alpha}$) are the number density (mass) of the proton and alpha, respectively.

Following the very recent studies \citep{amaro24,mostafavi24}, we use the Coulomb number ($N_{c}$) to analyze the collisional thermalization. $N_{c}$ denotes the approximation of collisions at the spacecraft position based on local plasma properties without taking the propagation effects of the CME and solar wind into account. The $N_{c}$ at the PSP location is calculated with Equations (5) and (6) \citep{mostafavi24}.

\begin{equation}
\rm N_{c}=R/(V_{p}\tau_{c})
\end{equation}

\begin{equation}
\rm \tau_{c}=133\frac{{(\omega _{\alpha p})^3}}{n_{p} };\
\rm \omega _{\alpha p}=\sqrt{\frac{2T_{\alpha}}{m_{\alpha}}+\frac{2T_{p}}{m_{p}}}
\end{equation}
where R is the distance between the Sun and PSP, $V_{p}$ is the solar wind speed and $\tau_{c}$ is the time scale for energy exchange between alphas and protons.

\section{Observations and Results}
To acquire a reliable measurement, the core of the solar wind distribution should be within the field of view (FOV) of SPAN-I. This usually occurs during encounters when PSP is close to the Sun and its lateral velocity is high enough \citep[e.g.,][]{mostafavi22}. In this case more solar wind ions flowing in the spacecraft frame can move into the ram-facing side of PSP.

When preparing this paper, the online catalog \citep{mostl17,mostl20} shows that three CMEs impacted PSP inside 0.1 au until 2023 October 3. We first visually examine whether the core ions of each CME are in the FOV of SPAN-I to ensure reliable plasma moments. The results show that one CME meets the criterion. We also check the data quality flags of SWEAP and FIELDs to ensure that the data are reliable in this study.

The CME occurred on 2022 June 2 (Encounter 12) when STEREO-A and PSP were located 28.4$^{\circ}$ and 62.5$^{\circ}$ east of the Earth, respectively, as shown in Figure 1(a). The two black dashed lines indicate the directions toward the Earth and STEREO-A as labelled. The purple dashed line describes the PSP trajectory. The blue arrow denotes the longitude of the active region (AR), which is the CME propagation direction in the ecliptic plane if no deflection.

The CME was observed by remote sensing instruments \citep{braga24}. It is associated with a filament eruption originating from an AR (named NOAA 13029 later) located at heliographic coordinate S17E100 from the Earth perspective. Thus the AIA can only observe the off-limb signatures of the filament as shown in Figure 1(b). The source region is well observed by STEREO-A. Figure 1(c) presents the CME front with the difference image of EUVI-A 195 \AA. Figure 1(d) displays the CME with the difference image of COR2-A, which intercepted PSP at $\sim$15 $R_\odot$ later \citep{braga24}.

Figure 2 presents the in situ measurements of the CME and surrounding solar wind from 11:00 to 15:00 UT on 2022 June 2, with two red vertical dashed lines in each panel demarcating two boundaries of the CME ejecta. Panels (a) and (b) illustrate the azimuthal fluxes of protons and alphas in the spacecraft frame, showing that the core ions are primarily in the FOV of SPAN-I. Panel (c) displays the total magnetic field and its three components in the RTN coordinate. No typical magnetic field rotation is observed within the ejecta and the ejecta duration is only about 2 hrs (from 11:51 to 14:08 UT), mainly because PSP encounters the CME flank rather than its apex \citep{braga24}.

The proton density ($n_{p}$) and the alpha-to-proton density ratio ($n_{\alpha}/n_{p}$) are presented in Panel (d), and the velocity of protons and alphas, in Panel (e). The CME velocity is slow thus no shock is driven ahead of it. Panel (f) displays the Alfv\'{e}n Mach number (M$_{A}$) and the plasma $\beta$, in which two blue horizontal dashed lines mark the levels of $M_{A}=1$ and $\beta=0.1$. The $\beta$ values of the CME are smaller compared to the surrounding solar wind, which is one characteristic to identify CMEs in the interplanetary space.

The last three panels display the $V_{\alpha p}$ and $V_{\alpha p}/V_{A}$, the proton temperature ($T_{p}$) and the alpha-to-proton temperature ratio ($T_{\alpha}/T_{p}$), as well as the Coulomb number ($N_{c}$) sequentially. The $V_{\alpha p}$ is obvious and generally lower than the local $V_{A}$. A very interesting point is that the ejecta duration can be divided into two intervals, labeled A and B, with $N_{c}=0.5$ as the dividing line, which is shown with black vertical dashed lines in Panels (g)--(i). Intervals A and B have different non-equilibrium features, which are demonstrated intuitively in Figure 3.

The left panels of Figure 3 show the distributions of the $V_{\alpha p}$ and $T_{\alpha}/T_{p}$ for both $N_{c} \leq 0.5$ (interval A) in green and $N_{c} > 0.5$ (interval B) in red. It is clear that the two intervals have different distribution ranges and average values for both the differential flow and temperature ratio. The average $V_{\alpha p}$ of intervals A and B are 96.52 and 21.96 km s$^{-1}$, respectively, which are nonnegligible differential flow. The average $T_{\alpha}/T_{p}$ of the two intervals are 7.65 and 2.23 sequentially.

The right panels of Figure 3 show the distributions of $V_{\alpha p}$ and $T_{\alpha}/T_{p}$ as a function of $N_{c}$. The vertical dashed lines in both panels denote the position of $N_{c}=0.5$. The horizontal dashed lines in Figures 3(b) and (d) delineate the $V_{\alpha p}=45$ km s$^{-1}$ and $T_{\alpha}/T_{p}=4$, respectively. The data points with $V_{\alpha p}>45$ km s$^{-1}$ or $T_{\alpha}/T_{p}>$ 4 mainly distribute in the space with $N_{c}<0.5$. Both the $V_{\alpha p}$ and $T_{\alpha}/T_{p}$ decreases as the $N_{c}$ increases, similar to observations near 1 au \citep{kasper08,kasper17}. This directly demonstrates that the non-equilibrium features of CMEs are correlated well with $N_{c}$, consistent with the situation of solar wind near the Sun \citep{mostafavi24}.

\section{Summary and Discussion}
In this Letter, we reported a CME that occurred on 2022 June 2 and was observed by both remote sensing instruments aboard SDO and STEREO-A near 1 au and in situ instruments aboard PSP at $\sim$15 $R_{\odot}$. The PSP observations demonstrated that obvious $V_{\alpha p}$ existed within the CME inside 0.1 au, which is different from the situation beyond 0.3 au. The $V_{\alpha p}$ of CMEs warns us that their absolute abundance of alphas might vary during transportation from the Sun to 0.3 au, similar to the slow solar wind.

Besides, the CME ejecta duration can be divided into intervals A and B with Coulomb number below and beyond 0.5, respectively. The average $V_{\alpha p}$ and $T_{\alpha}/T_{p}$ in interval A are 96.52 km s$^{-1}$ and 7.65, respectively, which are obviously larger compared with the corresponding values (21.96 km s$^{-1}$ and 2.23) of interval B. This directly illustrates that Coulomb collisions play an important role in reducing the non-equilibrium features.

As mentioned, there exits the preferential acceleration and heating of alphas and heavy ions near the Sun. \cite{kasper17} reported that the zone of preferential ion heating extends from the solar transition region to an outer boundary $\sim$0.1--0.2 au from the Sun. Combined the obvious $V_{\alpha p}$ of CMEs within 0.1 au (the current study) and the negligible $V_{\alpha p}$ of CMEs beyond 0.3 au \citep{liuying06,durovcova17}, it is reasonable to speculate that the alphas of CMEs are also accelerated preferentially within the zone, and the alpha-proton differential flow can be thermalized rapidly between 0.1 and 0.3 au by Coulomb collision \citep{kasper08} and/or alpha-proton instability \citep{gary00}.

\acknowledgments We thank the anonymous referee for the comments and suggestions that improved the original manuscript. We are grateful to Drs. Jia Huang (University of California, Berkeley), Bo Li (Shandong University), Jiansen He (Peking University), Pengfei Chen (Nanjing University), and Ying D. Liu (National Space Science Center, CAS) for their helpful discussions. This work is supported by the Strategic Priority Research Program of the Chinese Academy of Sciences (Grant No. XDB0560000), the NSFC grants 12373062, 11973031, and 12073042, as well as the National Key R\&D Program of China 2022YFF0503003 (2022YFF0503000). The authors acknowledge the use of data from the PSP, SDO, and STEREO.

\clearpage

\begin{figure*}[htb!]
\epsscale{1.0} \plotone{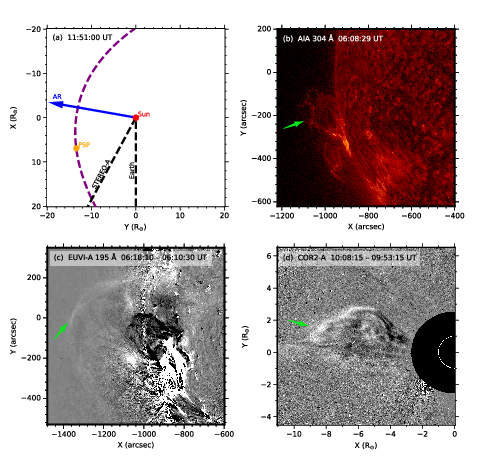} \caption{Remote sensing observations of the CME occurred on 2022 June 2. (a) Positions and/or directions of the Earth (SDO), STEREO-A, and PSP relative to the Sun in the ecliptic plane. (b) Direct image of AIA 304 \AA. (c) Difference image of EUVI-A 195 \AA. (d) Difference image of COR2-A. The green arrows denote the erupting filament or CME. \label{Figure 1}}
\end{figure*}

\begin{figure*}[htb!]
\epsscale{1.0} \plotone{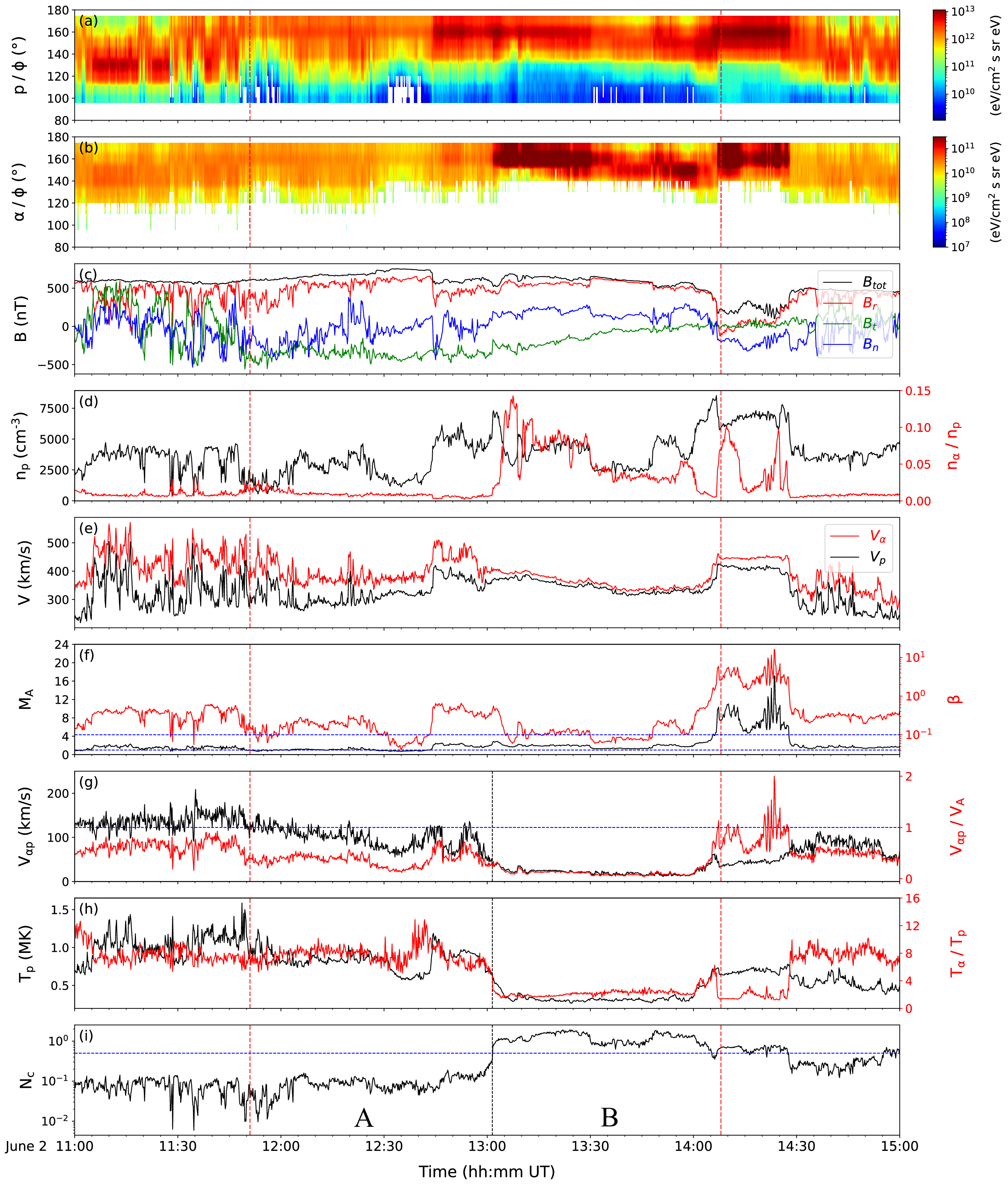} \caption{In situ measurements of the CME and solar wind by PSP from 11:00 to 15:00 UT on 2022 June 2. (a) and (b) The fluxes of protons and alphas as a function of azimuthal angle in the spacecraft frame. (c) Magnetic field. (d) The number density of protons and the absolute abundance of alphas. (e) The velocity of protons and alphas. (f) Alfv\'{e}n Mach number and plasma $\beta$. (g) Differential flow and the flow normalized to the local Alfv\'{e}n speed. (h) Temperature of protons and alpha-to-proton temperature ratio. (i) Coulomb number. \label{Figure 5}}
\end{figure*}

\begin{figure*}[htb!]
\epsscale{1.0} \plotone{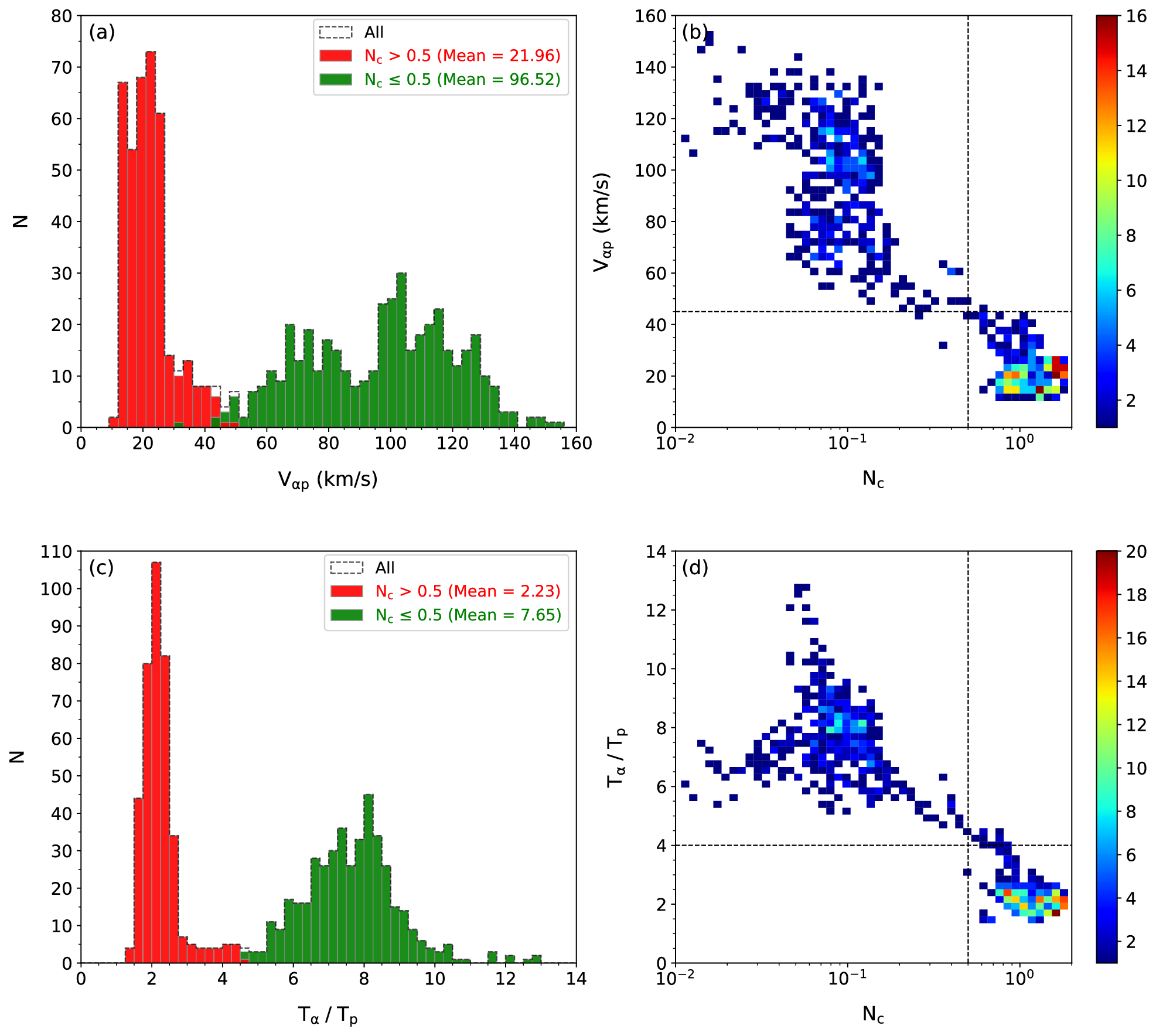} \caption{(a) and (c) Distributions of $V_{\alpha p}$ and $T_{\alpha}/T_{p}$ of the CME for different Coulomb number ranges. Dashed line, red, and green shaded regions show the $V_{\alpha p}$ (or $T_{\alpha}/T_{p}$) of the combined, Nc $\leq$ 0.5, and Nc $>$ 0.5, respectively. (b) and (d) Distributions of $V_{\alpha p}$ and $T_{\alpha}/T_{p}$ as a function of $N_{c}$. \label{Figure 6}}
\end{figure*}


\begin{thebibliography}{}
\expandafter\ifx\csname natexlab\endcsname\relax\def\natexlab#1{#1}\fi
\providecommand{\url}[1]{\href{#1}{#1}}
\providecommand{\dodoi}[1]{doi:~\href{http://doi.org/#1}{\nolinkurl{#1}}}
\providecommand{\doeprint}[1]{\href{http://ascl.net/#1}{\nolinkurl{http://ascl.net/#1}}}
\providecommand{\doarXiv}[1]{\href{https://arxiv.org/abs/#1}{\nolinkurl{https://arxiv.org/abs/#1}}}

\bibitem[{{Amaro} \& {Vaivads}(2024)}]{amaro24}
{Amaro}, M.~B., \& {Vaivads}, A. 2024, \apjl, 964, L2,
  \dodoi{10.3847/2041-8213/ad2ded}

\bibitem[{{Asbridge} {et~al.}(1976){Asbridge}, {Bame}, {Feldman}, \&
  {Montgomery}}]{asbridge76}
{Asbridge}, J.~R., {Bame}, S.~J., {Feldman}, W.~C., \& {Montgomery}, M.~D.
  1976, \jgr, 81, 2719, \dodoi{10.1029/JA081i016p02719}

\bibitem[{{Baker} {et~al.}(2013){Baker}, {Brooks}, {D{\'e}moulin}, {van
  Driel-Gesztelyi}, {Green}, {Steed}, \& {Carlyle}}]{baker13}
{Baker}, D., {Brooks}, D.~H., {D{\'e}moulin}, P., {et~al.} 2013, \apj, 778, 69,
  \dodoi{10.1088/0004-637X/778/1/69}

\bibitem[{{Bale} {et~al.}(2016){Bale}, {Goetz}, {Harvey}, {Turin}, {Bonnell},
  {Dudok de Wit}, {Ergun}, {MacDowall}, {Pulupa}, {Andre}, {Bolton},
  {Bougeret}, {Bowen}, {Burgess}, {Cattell}, {Chandran}, {Chaston}, {Chen},
  {Choi}, {Connerney}, {Cranmer}, {Diaz-Aguado}, {Donakowski}, {Drake},
  {Farrell}, {Fergeau}, {Fermin}, {Fischer}, {Fox}, {Glaser}, {Goldstein},
  {Gordon}, {Hanson}, {Harris}, {Hayes}, {Hinze}, {Hollweg}, {Horbury},
  {Howard}, {Hoxie}, {Jannet}, {Karlsson}, {Kasper}, {Kellogg}, {Kien},
  {Klimchuk}, {Krasnoselskikh}, {Krucker}, {Lynch}, {Maksimovic}, {Malaspina},
  {Marker}, {Martin}, {Martinez-Oliveros}, {McCauley}, {McComas}, {McDonald},
  {Meyer-Vernet}, {Moncuquet}, {Monson}, {Mozer}, {Murphy}, {Odom},
  {Oliverson}, {Olson}, {Parker}, {Pankow}, {Phan}, {Quataert}, {Quinn},
  {Ruplin}, {Salem}, {Seitz}, {Sheppard}, {Siy}, {Stevens}, {Summers}, {Szabo},
  {Timofeeva}, {Vaivads}, {Velli}, {Yehle}, {Werthimer}, \& {Wygant}}]{bale16}
{Bale}, S.~D., {Goetz}, K., {Harvey}, P.~R., {et~al.} 2016, \ssr, 204, 49,
  \dodoi{10.1007/s11214-016-0244-5}

\bibitem[{{Berger} {et~al.}(2011){Berger}, {Wimmer-Schweingruber}, \&
  {Gloeckler}}]{berger11}
{Berger}, L., {Wimmer-Schweingruber}, R.~F., \& {Gloeckler}, G. 2011, \prl,
  106, 151103, \dodoi{10.1103/PhysRevLett.106.151103}

\bibitem[{{Braga} {et~al.}(2024){Braga}, {Jagarlamudi}, {Vourlidas},
  {Stenborg}, \& {Nieves-Chinchilla}}]{braga24}
{Braga}, C.~R., {Jagarlamudi}, V.~K., {Vourlidas}, A., {Stenborg}, G., \&
  {Nieves-Chinchilla}, T. 2024, \apj, 965, 185,
  \dodoi{10.3847/1538-4357/ad2b4e}

\bibitem[{{Chandran}(2010)}]{chandran10}
{Chandran}, B. D.~G. 2010, \apj, 720, 548, \dodoi{10.1088/0004-637X/720/1/548}

\bibitem[{{Chen}(2011)}]{chenpengfei11}
{Chen}, P.~F. 2011, Living Reviews in Solar Physics, 8, 1,
  \dodoi{10.12942/lrsp-2011-1}

\bibitem[{{Cheng} {et~al.}(2020){Cheng}, {Zhang}, {Kliem}, {T{\"o}r{\"o}k},
  {Xing}, {Zhou}, {Inhester}, \& {Ding}}]{chengxin20}
{Cheng}, X., {Zhang}, J., {Kliem}, B., {et~al.} 2020, \apj, 894, 85,
  \dodoi{10.3847/1538-4357/ab886a}

\bibitem[{{Chhiber} {et~al.}(2016){Chhiber}, {Usmanov}, {Matthaeus}, \&
  {Goldstein}}]{chhiber16}
{Chhiber}, R., {Usmanov}, A., {Matthaeus}, W., \& {Goldstein}, M. 2016, \apj,
  821, 34, \dodoi{10.3847/0004-637X/821/1/34}

\bibitem[{{Formisano} {et~al.}(1970){Formisano}, {Palmiotto}, \&
  {Moreno}}]{formisano70}
{Formisano}, V., {Palmiotto}, F., \& {Moreno}, G. 1970, \solphys, 15, 479,
  \dodoi{10.1007/BF00151853}

\bibitem[{{Fox} {et~al.}(2016){Fox}, {Velli}, {Bale}, {Decker}, {Driesman},
  {Howard}, {Kasper}, {Kinnison}, {Kusterer}, {Lario}, {Lockwood}, {McComas},
  {Raouafi}, \& {Szabo}}]{fox16}
{Fox}, N.~J., {Velli}, M.~C., {Bale}, S.~D., {et~al.} 2016, \ssr, 204, 7,
  \dodoi{10.1007/s11214-015-0211-6}

\bibitem[{{Fu} {et~al.}(2018){Fu}, {Madjarska}, {Li}, {Xia}, \&
  {Huang}}]{fuhui18}
{Fu}, H., {Madjarska}, M.~S., {Li}, B., {Xia}, L., \& {Huang}, Z. 2018, \mnras,
  478, 1884, \dodoi{10.1093/mnras/sty1211}

\bibitem[{{Gary} {et~al.}(2000){Gary}, {Yin}, {Winske}, \&
  {Reisenfeld}}]{gary00}
{Gary}, S.~P., {Yin}, L., {Winske}, D., \& {Reisenfeld}, D.~B. 2000, \grl, 27,
  1355, \dodoi{10.1029/2000GL000019}

\bibitem[{{Gopalswamy} {et~al.}(2003){Gopalswamy}, {Shimojo}, {Lu}, {Yashiro},
  {Shibasaki}, \& {Howard}}]{gopalswamy03}
{Gopalswamy}, N., {Shimojo}, M., {Lu}, W., {et~al.} 2003, \apj, 586, 562,
  \dodoi{10.1086/367614}

\bibitem[{{Howard} {et~al.}(2008){Howard}, {Moses}, {Vourlidas}, {Newmark},
  {Socker}, {Plunkett}, {Korendyke}, {Cook}, {Hurley}, {Davila}, {Thompson},
  {St Cyr}, {Mentzell}, {Mehalick}, {Lemen}, {Wuelser}, {Duncan}, {Tarbell},
  {Wolfson}, {Moore}, {Harrison}, {Waltham}, {Lang}, {Davis}, {Eyles},
  {Mapson-Menard}, {Simnett}, {Halain}, {Defise}, {Mazy}, {Rochus}, {Mercier},
  {Ravet}, {Delmotte}, {Auchere}, {Delaboudiniere}, {Bothmer}, {Deutsch},
  {Wang}, {Rich}, {Cooper}, {Stephens}, {Maahs}, {Baugh}, {McMullin}, \&
  {Carter}}]{howard08}
{Howard}, R.~A., {Moses}, J.~D., {Vourlidas}, A., {et~al.} 2008, \ssr, 136, 67,
  \dodoi{10.1007/s11214-008-9341-4}

\bibitem[{{Kaiser}(2005)}]{kaiser05}
{Kaiser}, M.~L. 2005, Advances in Space Research, 36, 1483,
  \dodoi{10.1016/j.asr.2004.12.066}

\bibitem[{{Kasper} {et~al.}(2008){Kasper}, {Lazarus}, \& {Gary}}]{kasper08}
{Kasper}, J.~C., {Lazarus}, A.~J., \& {Gary}, S.~P. 2008, \prl, 101, 261103,
  \dodoi{10.1103/PhysRevLett.101.261103}

\bibitem[{{Kasper} {et~al.}(2013){Kasper}, {Maruca}, {Stevens}, \&
  {Zaslavsky}}]{kasper13}
{Kasper}, J.~C., {Maruca}, B.~A., {Stevens}, M.~L., \& {Zaslavsky}, A. 2013,
  \prl, 110, 091102, \dodoi{10.1103/PhysRevLett.110.091102}

\bibitem[{{Kasper} {et~al.}(2016){Kasper}, {Abiad}, {Austin}, {Balat-Pichelin},
  {Bale}, {Belcher}, {Berg}, {Bergner}, {Berthomier}, {Bookbinder}, {Brodu},
  {Caldwell}, {Case}, {Chandran}, {Cheimets}, {Cirtain}, {Cranmer}, {Curtis},
  {Daigneau}, {Dalton}, {Dasgupta}, {DeTomaso}, {Diaz-Aguado}, {Djordjevic},
  {Donaskowski}, {Effinger}, {Florinski}, {Fox}, {Freeman}, {Gallagher},
  {Gary}, {Gauron}, {Gates}, {Goldstein}, {Golub}, {Gordon}, {Gurnee}, {Guth},
  {Halekas}, {Hatch}, {Heerikuisen}, {Ho}, {Hu}, {Johnson}, {Jordan},
  {Korreck}, {Larson}, {Lazarus}, {Li}, {Livi}, {Ludlam}, {Maksimovic},
  {McFadden}, {Marchant}, {Maruca}, {McComas}, {Messina}, {Mercer}, {Park},
  {Peddie}, {Pogorelov}, {Reinhart}, {Richardson}, {Robinson}, {Rosen},
  {Skoug}, {Slagle}, {Steinberg}, {Stevens}, {Szabo}, {Taylor}, {Tiu}, {Turin},
  {Velli}, {Webb}, {Whittlesey}, {Wright}, {Wu}, \& {Zank}}]{kasper16}
{Kasper}, J.~C., {Abiad}, R., {Austin}, G., {et~al.} 2016, \ssr, 204, 131,
  \dodoi{10.1007/s11214-015-0206-3}

\bibitem[{{Kasper} {et~al.}(2017){Kasper}, {Klein}, {Weber}, {Maksimovic},
  {Zaslavsky}, {Bale}, {Maruca}, {Stevens}, \& {Case}}]{kasper17}
{Kasper}, J.~C., {Klein}, K.~G., {Weber}, T., {et~al.} 2017, \apj, 849, 126,
  \dodoi{10.3847/1538-4357/aa84b1}

\bibitem[{{Lemen} {et~al.}(2012){Lemen}, {Title}, {Akin}, {Boerner}, {Chou},
  {Drake}, {Duncan}, {Edwards}, {Friedlaender}, {Heyman}, {Hurlburt}, {Katz},
  {Kushner}, {Levay}, {Lindgren}, {Mathur}, {McFeaters}, {Mitchell}, {Rehse},
  {Schrijver}, {Springer}, {Stern}, {Tarbell}, {Wuelser}, {Wolfson}, {Yanari},
  {Bookbinder}, {Cheimets}, {Caldwell}, {Deluca}, {Gates}, {Golub}, {Park},
  {Podgorski}, {Bush}, {Scherrer}, {Gummin}, {Smith}, {Auker}, {Jerram},
  {Pool}, {Soufli}, {Windt}, {Beardsley}, {Clapp}, {Lang}, \&
  {Waltham}}]{lemen12}
{Lemen}, J.~R., {Title}, A.~M., {Akin}, D.~J., {et~al.} 2012, \solphys, 275,
  17, \dodoi{10.1007/s11207-011-9776-8}

\bibitem[{{Lepri} \& {Rivera}(2021)}]{lepri21}
{Lepri}, S.~T., \& {Rivera}, Y.~J. 2021, \apj, 912, 51,
  \dodoi{10.3847/1538-4357/abea9f}

\bibitem[{{Liu} {et~al.}(2006){Liu}, {Richardson}, {Belcher}, {Kasper}, \&
  {Elliott}}]{liuying06}
{Liu}, Y., {Richardson}, J.~D., {Belcher}, J.~W., {Kasper}, J.~C., \&
  {Elliott}, H.~A. 2006, Journal of Geophysical Research (Space Physics), 111,
  A01102, \dodoi{10.1029/2005JA011329}

\bibitem[{{Marsch} {et~al.}(1982){Marsch}, {Rosenbauer}, {Schwenn},
  {Muehlhaeuser}, \& {Neubauer}}]{marsch82b}
{Marsch}, E., {Rosenbauer}, H., {Schwenn}, R., {Muehlhaeuser}, K.~H., \&
  {Neubauer}, F.~M. 1982, \jgr, 87, 35, \dodoi{10.1029/JA087iA01p00035}

\bibitem[{{Maruca} {et~al.}(2013){Maruca}, {Bale}, {Sorriso-Valvo}, {Kasper},
  \& {Stevens}}]{maruca13}
{Maruca}, B.~A., {Bale}, S.~D., {Sorriso-Valvo}, L., {Kasper}, J.~C., \&
  {Stevens}, M.~L. 2013, \prl, 111, 241101,
  \dodoi{10.1103/PhysRevLett.111.241101}

\bibitem[{{Mostafavi} {et~al.}(2022){Mostafavi}, {Allen}, {McManus}, {Ho},
  {Raouafi}, {Larson}, {Kasper}, \& {Bale}}]{mostafavi22}
{Mostafavi}, P., {Allen}, R.~C., {McManus}, M.~D., {et~al.} 2022, \apjl, 926,
  L38, \dodoi{10.3847/2041-8213/ac51e1}

\bibitem[{{Mostafavi} {et~al.}(2024){Mostafavi}, {Allen}, {Jagarlamudi},
  {Bourouaine}, {Badman}, {Ho}, {Raouafi}, {Hill}, {Verniero}, {Larson},
  {Kasper}, \& {Bale}}]{mostafavi24}
{Mostafavi}, P., {Allen}, R.~C., {Jagarlamudi}, V.~K., {et~al.} 2024, \aap,
  682, A152, \dodoi{10.1051/0004-6361/202347134}

\bibitem[{{M{\"o}stl} {et~al.}(2017){M{\"o}stl}, {Isavnin}, {Boakes}, {Kilpua},
  {Davies}, {Harrison}, {Barnes}, {Krupar}, {Eastwood}, {Good}, {Forsyth},
  {Bothmer}, {Reiss}, {Amerstorfer}, {Winslow}, {Anderson}, {Philpott},
  {Rodriguez}, {Rouillard}, {Gallagher}, {Nieves-Chinchilla}, \&
  {Zhang}}]{mostl17}
{M{\"o}stl}, C., {Isavnin}, A., {Boakes}, P.~D., {et~al.} 2017, Space Weather,
  15, 955, \dodoi{10.1002/2017SW001614}

\bibitem[{{M{\"o}stl} {et~al.}(2020){M{\"o}stl}, {Weiss}, {Bailey}, {Reiss},
  {Amerstorfer}, {Hinterreiter}, {Bauer}, {McIntosh}, {Lugaz}, \&
  {Stansby}}]{mostl20}
{M{\"o}stl}, C., {Weiss}, A.~J., {Bailey}, R.~L., {et~al.} 2020, \apj, 903, 92,
  \dodoi{10.3847/1538-4357/abb9a1}

\bibitem[{{Neugebauer} {et~al.}(1994){Neugebauer}, {Goldstein}, {Bame}, \&
  {Feldman}}]{neugebauer94}
{Neugebauer}, M., {Goldstein}, B.~E., {Bame}, S.~J., \& {Feldman}, W.~C. 1994,
  \jgr, 99, 2505, \dodoi{10.1029/93JA02615}

\bibitem[{{Neugebauer} {et~al.}(1996){Neugebauer}, {Goldstein}, {Smith}, \&
  {Feldman}}]{neugebauer96}
{Neugebauer}, M., {Goldstein}, B.~E., {Smith}, E.~J., \& {Feldman}, W.~C. 1996,
  \jgr, 101, 17047, \dodoi{10.1029/96JA01406}

\bibitem[{{Pesnell} {et~al.}(2012){Pesnell}, {Thompson}, \&
  {Chamberlin}}]{pesnell12}
{Pesnell}, W.~D., {Thompson}, B.~J., \& {Chamberlin}, P.~C. 2012, \solphys,
  275, 3, \dodoi{10.1007/s11207-011-9841-3}

\bibitem[{{Ran} {et~al.}(2024){Ran}, {Liu}, {Chen}, \& {Mostafavi}}]{ranhao24}
{Ran}, H., {Liu}, Y.~D., {Chen}, C., \& {Mostafavi}, P. 2024, \apj, 963, 82,
  \dodoi{10.3847/1538-4357/ad2069}

\bibitem[{{Reisenfeld} {et~al.}(2001){Reisenfeld}, {Gary}, {Gosling},
  {Steinberg}, {McComas}, {Goldstein}, \& {Neugebauer}}]{reisenfeld01}
{Reisenfeld}, D.~B., {Gary}, S.~P., {Gosling}, J.~T., {et~al.} 2001, \jgr, 106,
  5693, \dodoi{10.1029/2000JA000317}

\bibitem[{{Robbins} {et~al.}(1970){Robbins}, {Hundhausen}, \&
  {Bame}}]{robbins70}
{Robbins}, D.~E., {Hundhausen}, A.~J., \& {Bame}, S.~J. 1970, \jgr, 75, 1178,
  \dodoi{10.1029/JA075i007p01178}

\bibitem[{{Song} \& {Yao}(2020)}]{song20b}
{Song}, H., \& {Yao}, S. 2020, Sci China Tech Sci, 63, 2171,
  \dodoi{10.1007/s11431-020-1680-y}

\bibitem[{{Song} {et~al.}(2018){Song}, {Chen}, {Qiu}, {Chen}, {Zhang}, {Cheng},
  {Shen}, \& {Zheng}}]{song18a}
{Song}, H.~Q., {Chen}, Y., {Qiu}, J., {et~al.} 2018, \apjl, 857, L21,
  \dodoi{10.3847/2041-8213/aabcc3}

\bibitem[{{Song} {et~al.}(2013){Song}, {Chen}, {Ye}, {Han}, {Du}, {Li},
  {Zhang}, \& {Hu}}]{song13}
{Song}, H.~Q., {Chen}, Y., {Ye}, D.~D., {et~al.} 2013, \apj, 773, 129,
  \dodoi{10.1088/0004-637X/773/2/129}

\bibitem[{{Song} {et~al.}(2016){Song}, {Zhong}, {Chen}, {Zhang}, {Cheng},
  {Zhao}, {Hu}, \& {Li}}]{song16}
{Song}, H.~Q., {Zhong}, Z., {Chen}, Y., {et~al.} 2016, \apjs, 224, 27,
  \dodoi{10.3847/0067-0049/224/2/27}

\bibitem[{{Song} {et~al.}(2017){Song}, {Chen}, {Li}, {Li}, {Zhao}, {He},
  {Duan}, {Cheng}, \& {Zhang}}]{song17a}
{Song}, H.~Q., {Chen}, Y., {Li}, B., {et~al.} 2017, \apjl, 836, L11,
  \dodoi{10.3847/2041-8213/aa5d54}

\bibitem[{{{\v{D}}urovcov{\'a}} {et~al.}(2017){{\v{D}}urovcov{\'a}},
  {{\v{S}}afr{\'a}nkov{\'a}}, {N{\v{e}}me{\v{c}}ek}, \&
  {Richardson}}]{durovcova17}
{{\v{D}}urovcov{\'a}}, T., {{\v{S}}afr{\'a}nkov{\'a}}, J.,
  {N{\v{e}}me{\v{c}}ek}, Z., \& {Richardson}, J.~D. 2017, \apj, 850, 164,
  \dodoi{10.3847/1538-4357/aa9618}

\bibitem[{{von Steiger} {et~al.}(1995){von Steiger}, {Geiss}, {Gloeckler}, \&
  {Galvin}}]{vonSteiger95}
{von Steiger}, R., {Geiss}, J., {Gloeckler}, G., \& {Galvin}, A.~B. 1995, \ssr,
  72, 71, \dodoi{10.1007/BF00768756}

\bibitem[{{von Steiger} \& {Zurbuchen}(2006)}]{vonSteiger06}
{von Steiger}, R., \& {Zurbuchen}, T.~H. 2006, \grl, 33, L09103,
  \dodoi{10.1029/2005GL024998}

\bibitem[{{Wang} {et~al.}(2017){Wang}, {Liu}, {Wang}, {Hu}, {Shen}, {Jiang}, \&
  {Zhu}}]{wangwensi17}
{Wang}, W., {Liu}, R., {Wang}, Y., {et~al.} 2017, Nature Communications, 8,
  1330, \dodoi{10.1038/s41467-017-01207-x}

\bibitem[{{Wang} {et~al.}(2023){Wang}, {Song}, {Chen}, {Li}, {Hou}, \&
  {Zheng}}]{wangxinyue23}
{Wang}, X., {Song}, H., {Chen}, Y., {et~al.} 2023, \apj, 957, 58,
  \dodoi{10.3847/1538-4357/acff5d}

\bibitem[{{Webb} \& {Howard}(2012)}]{webb12}
{Webb}, D.~F., \& {Howard}, T.~A. 2012, Living Reviews in Solar Physics, 9, 3,
  \dodoi{10.12942/lrsp-2012-3}

\bibitem[{{Widing} \& {Feldman}(2001)}]{widing01}
{Widing}, K.~G., \& {Feldman}, U. 2001, \apj, 555, 426, \dodoi{10.1086/321482}

\bibitem[{{Zhang} {et~al.}(2024){Zhang}, {Song}, {Zhang}, {Fu}, {Li}, {Li},
  {Wang}, {Wang}, \& {Chen}}]{zhangxuechao24}
{Zhang}, X., {Song}, H., {Zhang}, C., {et~al.} 2024, \apj, 967, 118,
  \dodoi{10.3847/1538-4357/ad46f7}

\end{thebibliography}
\end{document}